\documentclass[useAMS,usenatbib]{mn2e} 
\input{epsf}

\newif\ifAMStwofonts
\AMStwofontstrue

\usepackage{url}
\usepackage{graphicx}

\makeatletter
\newlength{\abovecaptionskip}
\makeatother
\usepackage{threeparttable}

\def\lesssim{\mathrel{\hbox{\rlap{\hbox{\lower4pt\hbox{$\sim$}}}\hbox{$<$}}}}
\def\gtrsim{\mathrel{\hbox{\rlap{\hbox{\lower4pt\hbox{$\sim$}}}\hbox{$>$}}}}

\def\l_lsun{$\log{L/\rm L_{\odot}}$~}
\def\masa_msun{$M/ \rm M_{\odot}$~}

\def\m_mstar{$M/M_{*}$~}

\title[Formation of redback pulsars]{Identifying the formation mechanism 
of redback pulsars}

\author[M. A. De Vito, O. G. Benvenuto \& J. E. Horvath]
{
M. A. De  Vito$^{1,2}$\thanks{Member of  the Carrera del Investigador
Cient\'{\i}fico, Consejo Nacional de Investigaciones Cient\'\i ficas y
T\'ecnicas (CONICET). Email: adevito@fcaglp.unlp.edu.ar},
O. G. Benvenuto$^{1,2}$\thanks{Member of  the Carrera del Investigador
Cient\'{\i}fico, Comisi\'on de  Investigaciones Cient\'{\i}ficas de la  
Provincia de Buenos Aires (CIC). Email: obenvenu@fcaglp.unlp.edu.ar},
J. E. Horvath$^{3}$\thanks{Email: foton@iag.usp.br},
\\
$^{1}$ Instituto de Astrof\'{\i}sica de La Plata, IALP, CCT-CONICET-UNLP,
Argentina\\
$^{2}$ Facultad de Ciencias Astron\'omicas y Geof\'{\i}sicas, Universidad
Nacional de La Plata (UNLP),\\ Paseo del Bosque S/N, B1900FWA, La Plata,
Argentina\\
$^{3}$ Instituto de Astronom\'{\i}a, Geof\'{\i}sica e Ci\^encias
Atmosf\'ericas, Universidade de S\~ao Paulo,\\ R. do Mat\~ao 1226 (05508-090),
Cidade Universit\'aria, S\~ao Paulo SP, Brazil}

\begin{document}

\date{February 7, 2020}

\pagerange{\pageref{firstpage}--\pageref{lastpage}} \pubyear{2020}

\maketitle \label{firstpage}

\begin{abstract}

We analyse the evolution of close binary systems containing a neutron star that
lead to the formation of redback pulsars. Recently there has been some debate on
the origin of such systems and the formation mechanism of redbacks may still be
considered as an open problem.
We show that the operation of a strong evaporation mechanism, starting from the
moment when the donor star becomes fully convective (or alternatively since the
formation of the neutron star by accretion induced collapse), produces  systems
with donor masses and orbital periods in the range corresponding to redbacks
with donors appreciably smaller than their Roche lobes, i.e., they have
low filling factors (lower than $0.75$).
Models of redback pulsars can be constructed assuming the occurrence of
irradiation feedback. They have been shown to undergo cyclic mass transfer
during the epoch at which they attain donor masses and orbital periods
corresponding to redbacks, and stay in quasi-Roche lobe overflow
conditions with {\it high} filling factors.  We show that, if irradiation
feedback occurs and radio ejection inhibits further accretion onto the neutron
star after the first mass transfer cycle, the redback systems feature {\it high}
filling factors.
We suggest that the filling factor should be considered as a useful tool
for discriminating among those redback formation mechanisms. We compare
theoretical results with available observations, and conclude that observations
tend to favour models with high filling factors.

\end{abstract}


\section{Introduction} \label{sec:intro}

From the study of eclipsing millisecond pulsars (MSPs) belonging to close binary
systems (CBSs), evidence has emerged for the existence of two well-separated
families of systems:  Black Widows (BWs) and Redbacks (RBs)
\citep{2013IAUS..291..127R}. While RBs have circular orbits with orbital periods
$0.1 \lesssim P /d \lesssim 1.0$ and companion stars with masses
$M_{2}$ in the range $0.1 \lesssim M_{2}/M_{\odot}  \lesssim 0.7$, BWs show
orbital periods in the same range but substantially {\it lighter} companions,
with $M_{2}/M_{\odot} \lesssim 0.05$. Besides the range of characteristic
masses, it is the state of these companions which differentiates both
families. BWs companions are degenerate stars,  while RBs companions are normal
stars, with spectral types in a wide range, from F to M.  In addition, for the
same effective temperature, donor stars in RBs are brighter than a isolated main
sequence star. Therefore, the donor star in a RB system has a radius greater
than an isolated star of the same spectral type. For this reason, it is
said that donor stars in RB systems are extended stars.

It is important to understand the mechanisms that give rise to the  formation of
BWs and RBs. Their very existence challenges the standard  treatment
(\citealt{2002ApJ...565.1107P}; \citealt{2005MNRAS.362..891B}) of the evolution
of CBSs including a  neutron star (NS) component. For the case of BWs, a general
consensus emerged  that they stem from evaporation of the donor star driven by
pulsar irradiation  \citep{1988Natur.333..832P}. On the other hand, the
mechanism(s) for the formation of RBs remained controversial. To date there are
three mechanisms proposed to account for the existence of RBs: strong
evaporation, irradiation  feedback, and accretion induced collapse of white
dwarfs (WDs).

A strong evaporation mechanism has been presented in
\citet{2013ApJ...775...27C}. These authors considered that the pulsar 
begins  to irradiate when the donor star becomes fully convective. At that 
moment, magnetic braking ceases (it is no longer an angular momentum sink) and
the donor  detaches from its Roche lobe. Since then on, radio ejection
\citep{2001ApJ...560L..71B} starts to inhibit further accretion onto the NS and
forces evaporation. Then, the system evolves to longer orbital periods in the
range corresponding to RBs.

Irradiation feedback has been studied by \citet{2004A&A...423..281B} and applied
to RBs by \citet{2014ApJ...786L...7B}. When the donor star undergoes Roche lobe
overflow (RLOF), it starts to transfer mass to the NS. If accretion occurs, the
material falling onto the NS leads to the emission of X-ray radiation that
illuminates the donor star. If the donor has a thick enough outer convective
zone, irradiation may strongly affect its evolution. In many cases this makes
the CBS to undergo cyclic mass transfer \citep{2004A&A...423..281B}. This may
occur if  radio ejection does not suppress accretion. In between these cycles,
mass  transfer ceases and the NS may act as a pulsar. 

\citet{2015MNRAS.446.2540S} proposed that RBs may be formed by accretion
induced collapse (AIC) of heavy oxygen, neon and magnesium (ONeMg) WDs. This
path to form a RB system contemplates a binary system initially composed of a
star of $8-11~M_{\odot}$ and a companion of $\approx 1~M_{\odot}$. After the
RLOF of the more massive star, a common envelope stage and envelope ejection,
the system results in a ONeMg WD (that initially was the more massive star of
the pair) and its companion, that will fill its Roche lobe and start
transferring matter to the WD. When the WD reaches the limit mass value of
$1.37~M_{\odot}$, it undergoes accretion induced collapse to form a NS.
Eventually the system becomes detached and since then on, it is considered
to follow an evolution similar to that studied by \citet{2013ApJ...775...27C}.

Still, there is room for another formation mechanism: we suggest that if
irradiation feedback forces an early detachment from its Roche lobe and mass
transfer stops, then pulsar spin-down irradiation may start. If such irradiation
inhibits further accretion onto the NS, it is possible to form RBs even with low
evaporation rates. In this scenario it is not necessary to wait for the donor to
achieve a mass low enough to become fully convective. Thus, much more massive
donors may undergo evaporation and are candidates to become RBs.

Following  \citet{2014A&A...564A...1B}, let us define the filling factor as $FF=
R_{\rm 2}/R_{\rm RL}$ where $R_{\rm 2}$ and $R_{\rm RL}$ are the radius of the
donor star and its corresponding Roche lobe \citep{1983ApJ...268..368E}, 
respectively. It will be shown below that the mechanisms that invoke strong
evaporation generally lead to the occurrence of donor stars with sizes much
smaller than those of their associated Roche lobes. On the contrary, the
mechanisms that involve irradiation feedback give rise to donor stars in the
quasi-RLOF  state, with their lobes almost fully filled. It is the goal of this
paper to explore if filling factors may help us to discriminate which is the
most probable formation mechanism of RB pulsars.

The remainder of this paper is organized as follows. In
Section~\ref{sec:irradia_evapora} we describe the treatment of irradiation
feedback and evaporation we shall consider in the numerical calculations
presented in Section~\ref{sec:calcu}. In Section~\ref{sec:observaciones} we
present  the observational data currently available for RB systems and compare
it with theoretical calculations. Finally, in Section~\ref{sec:conclu} we
discuss the relevance of our results and give some concluding remarks. 

\section{Considering Irradiation Feedback and Evaporation}
\label{sec:irradia_evapora}

The calculations to be presented below have been performed with our binary
stellar evolution code, described in \citet{2003MNRAS.342...50B} and
\citet{2014ApJ...786L...7B}. We shall make a brief summary of the treatment
of irradiation feedback and evaporation included in it.

In order to consider irradiation feedback, we assume that the NS acts as a point
source releasing an accretion luminosity $L_{\rm acc}= G M_{1}
\dot{M}_{1}/R_{1}$, where $M_{1}$, $\dot{M}_{1}$, and $R_{1}$ are the mass,
accretion rate and radius of the NS, respectively. For isotropic emission, the
energy flux incident on the donor star that effectively participates in
the irradiation feedback process is $F_{\rm irr}= \alpha_{\rm irrad} L_{\rm
acc}/(4\pi a^{2})$, where $a$ is the orbital separation, and $\alpha_{\rm
irrad}$ is considered as a free parameter.

As usual, we shall describe the rate of evaporation of the donor star
$\dot{M}_{\rm 2,evap}$ with the simple prescription given by
\citet{1992MNRAS.254P..19S}:

\begin{equation}
\dot{M}_{\rm 2,evap}= - \frac{\alpha_{\rm evap}}{2 v^{2}_{\rm esc}} L_{\rm PSR}
\bigg( \frac{R_{2}}{a} \bigg)^{2}.
\end{equation}

\noindent Here, $v_{\rm esc}$ is the escape velocity from the donor star
surface, $L_{\rm PSR}$ is the pulsar luminosity, and $\alpha_{\rm evap}$ is a
free parameter kept constant in our simulations.

In order to compute the luminosity of the pulsar $L_{\rm PSR}$ given by $L_{\rm
PSR}=4\pi^{2} I \dot{P}_{\rm spin} P^{-3}_{\rm spin}$ (where $I$, $P_{\rm
spin}$, and $\dot{P}_{\rm spin}$ are the NS moment of inertia, spin period, and
its derivative, respectively) we make the same assumptions, as in
\citet{2013ApJ...775...27C}: $I= 10^{45}\ g\ cm^{2}$, initial $P_{\rm
spin}=3$~msec, $\dot{P}_{\rm spin}=10^{-20} s\ s^{-1}$, and a braking index
$n=3$. The evolution of the spin is computed considering the time since the
pulsar emission starts out.

As usual, we assumed that the NS accretes mass with a rate given by
$\dot{M}_{\rm NS}=min\big[\beta \dot{M}_{2}, \dot{M}_{\rm Edd}\big]$, where
$\beta$ is the efficiency of accretion, and $\dot{M}_{\rm Edd}$ is the critical
Eddington rate that represents the upper value possible for a NS. The material
lost from the system is assumed to carry the specific angular momentum of the
NS.

\section{Numerical Results} \label{sec:calcu}

Let us ignore irradiation feedback and compute the evolution of a CBS with  
Solar composition $1, 2$, and $3~M_{\odot}$ donor stars, together with a 
$1.4~M_{\odot}$ NS in tight orbits with initial periods  $P = 0.30,
0.60$, and $0.60$~d. We shall consider different values for the parameter
$\alpha_{\rm evap}=  0.010, 0.030, 0.075, 0.100,$ and $1.000$ (from moderate to
strong evaporation regimes) and assume that the pulsar starts to irradiate when
the donor becomes fully convective, inhibiting further accretion onto the NS.
This is essentially the exploration performed by \citet{2013ApJ...775...27C} for
the case of a $1~M_{\odot}$ donor. The results are shown in
Fig.~\ref{Fig:figure1} together with observational data on RBs presented below,
in Table~\ref{table:FF} and  BWs taken from from Patruno's
Catalogue\footnote{\url{https://apatruno.wordpress.com/about/millisecond-pulsar-catalogue/}}.

In our calculations, the donor stars with initial masses of $1, 2$, and
$3~M_{\odot}$ become fully convective when they have masses of $0.32~M_{\odot}$,
$0.14~M_{\odot}$, and $0.24~M_{\odot}$ at ages of $2.21, 3.79$, and $2.12$~Gyr,
respectively. Since then on, evaporation leads to a departure from the
predictions of ``standard'' CBS evolution.

Strong evaporation leads these systems to evolve to RB conditions, whereas
moderate evaporation produce BWs. Within this scenario it seems difficult to
populate the entire RB region indicated in the lower panel of
Fig.~\ref{Fig:figure1}. The reason is that this kind of models can only populate
the region of masses lower than the one it has when becomes fully convective
($M_{2} \lesssim 0.32~M_{\odot}$). This is appreciably lower than the
high-mass edge of the RB region usually considered in the literature. 
Moreover, stellar models constructed this way lead to the occurrence of 
high filling factors ($FF \gtrsim 0.75$) only shortly after detachment (less
than $1$~Gyr in the cases of $1$ and $3~M_{\odot}$, and in an even more
restricted  time interval for the case of $2~M_{\odot}$). The tracks
corresponding to $\alpha_{evap}=0.075$ can be considered as RB for a longer
period.  However, notice that they can populate a marginal portion of the plane
shown in Fig.~\ref{Fig:figure1}, and only for the cases of $1~M_{\odot}$ and
$3~M_{\odot}$. For the rest of the tracks, the time spent by these systems as
RBs with high filling factors is remarkably short. This can be seen
in  Fig.~\ref{Fig:FFvsT}, where we show the evolution of the filling factors 
for the same calculations presented in  Fig.~\ref{Fig:figure1}.

\begin{figure*}
\includegraphics[width=5.5in,angle=-90]{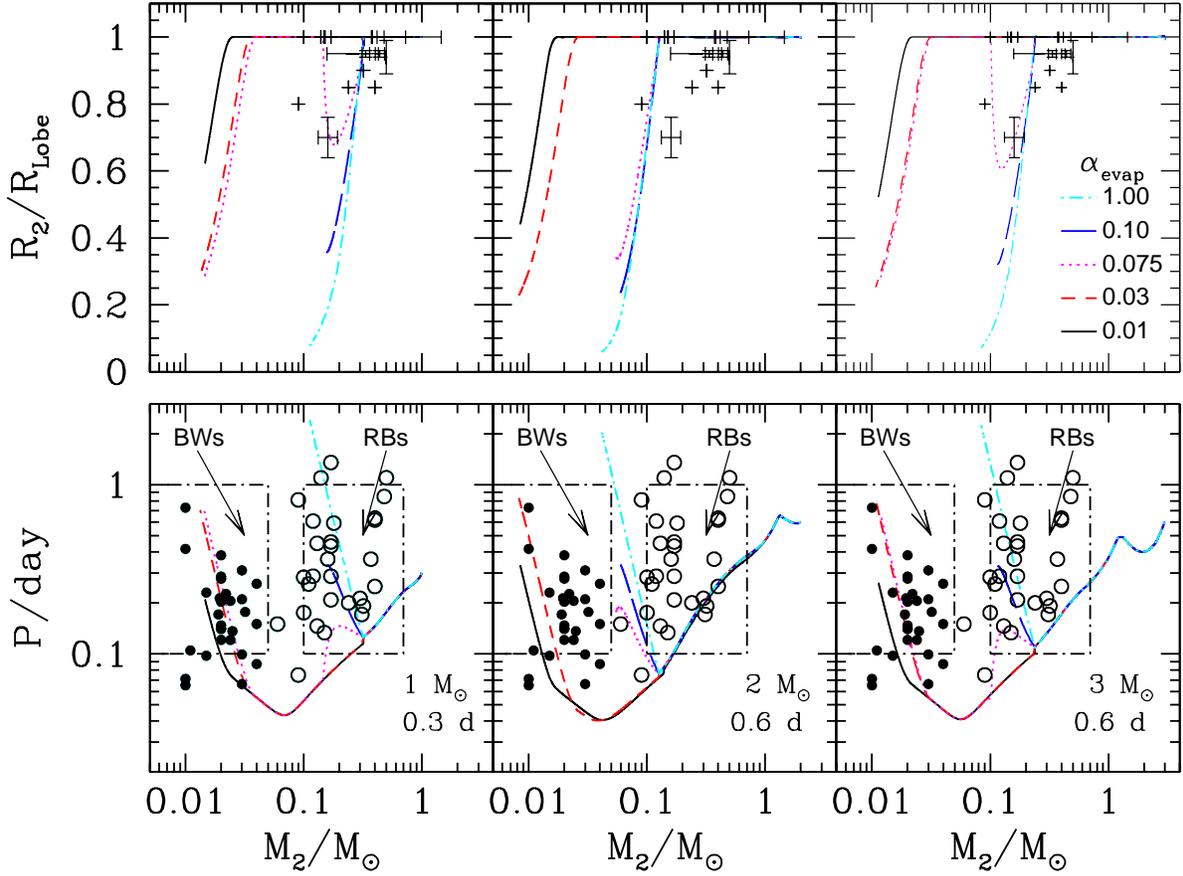}
\caption{The evolution of CBSs with donor stars of 1, 2, and 3~$M_{\odot}$
subject to different strengths of evaporation, switched on when they become
fully convective. Lower panel shows the evolution of the orbital period as a
function of donor mass. Strong evaporation regimes (corresponding to the highest
considered values of $\alpha_{evap}$) allow to reach orbital periods
corresponding to RBs. Upper panel shows the ratio of the donor radius to that of
the equivalent Roche lobe (i.e., the filling factor $FF$) for the {\it same set}
of evolutionary tracks. We have also plotted the $FF$ available in
Table~\ref{table:FF}. Notice that strong evaporation allows reaching orbital
periods in the whole range corresponding to RBs, but with {\textit low} filling factors.
Observed RBs listed in Table~\ref{table:FF} are represented with open circles,
whereas  BWs are shown with solid circles. \label{Fig:figure1}}
\end{figure*}

\begin{figure*}
\includegraphics[width=4.0in,angle=-90]{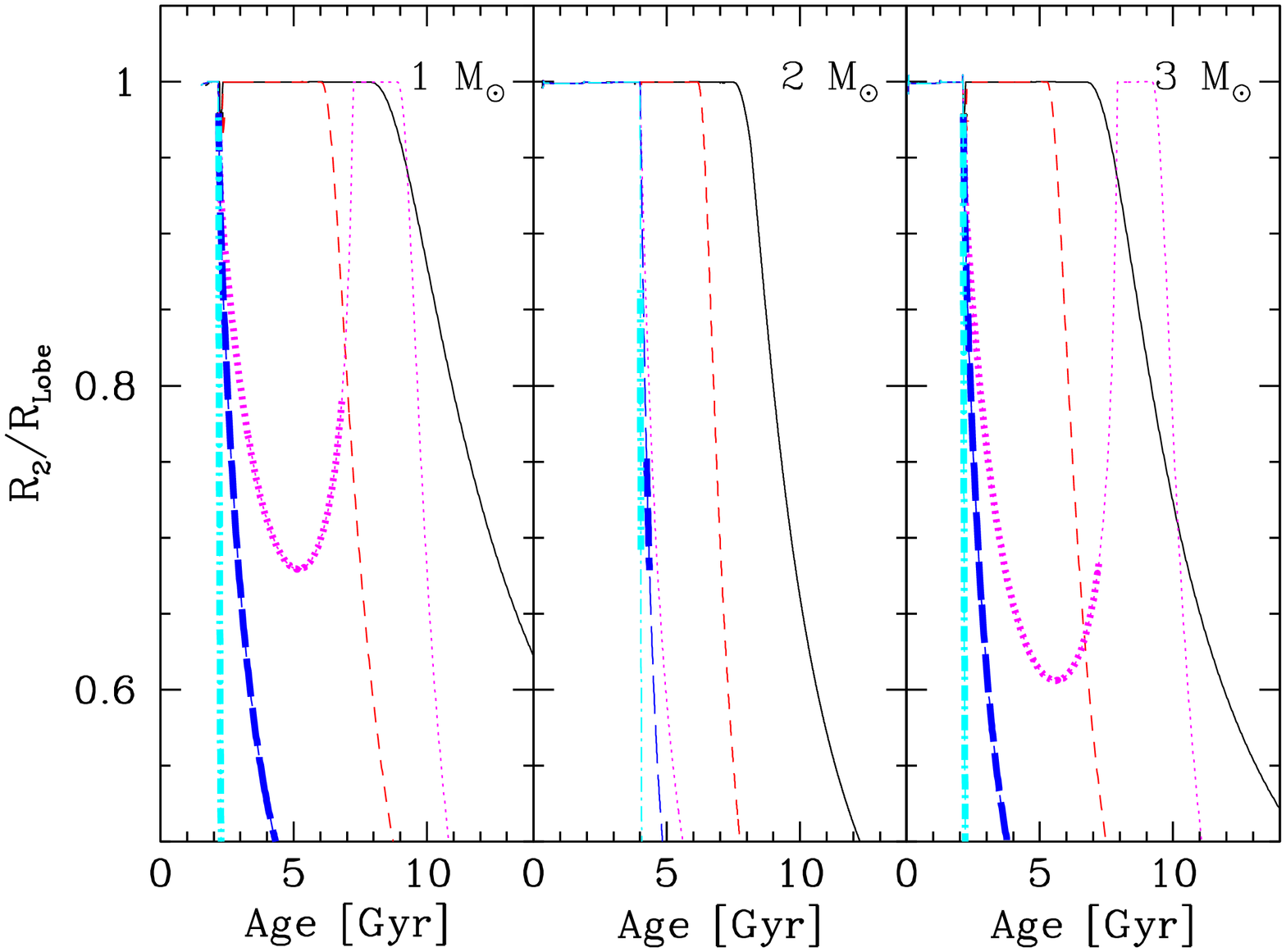}
\caption{The filling factor ($FF= R_{\rm 2}/R_{\rm RL}$) as a function of time 
for the evolutionary calculations presented in Fig.~\ref{Fig:figure1}. 
Here we employ the same type and colour of lines as in that Figure. 
We denote the stages when the system fulfils the conditions on its mass and 
orbital period to be considered as a RB
with heavy lines. \label{Fig:FFvsT}}
\end{figure*}

The models that consider irradiation feedback and do not impose any restriction
to the accretion onto the NS component of the pair have been presented in
Benvenuto, De Vito, \&  Horvath (2014; 2015; 2017). For appropriate initial
conditions, these models undergo a sequence of cyclic mass transfer episodes. In
each of these cycles, there is a short mass transfer episode followed by a long
period in which the donor remains slightly detached from its Roche lobe, without
shrinking back appreciably.  Because of this reason we have called it as a
``quasi-RLOF'' and $FF \gtrsim 0.90$. For
further details of this scenario we refer the reader to the above cited papers.

\begin{figure*}
\includegraphics[width=5.5in,angle=-90]{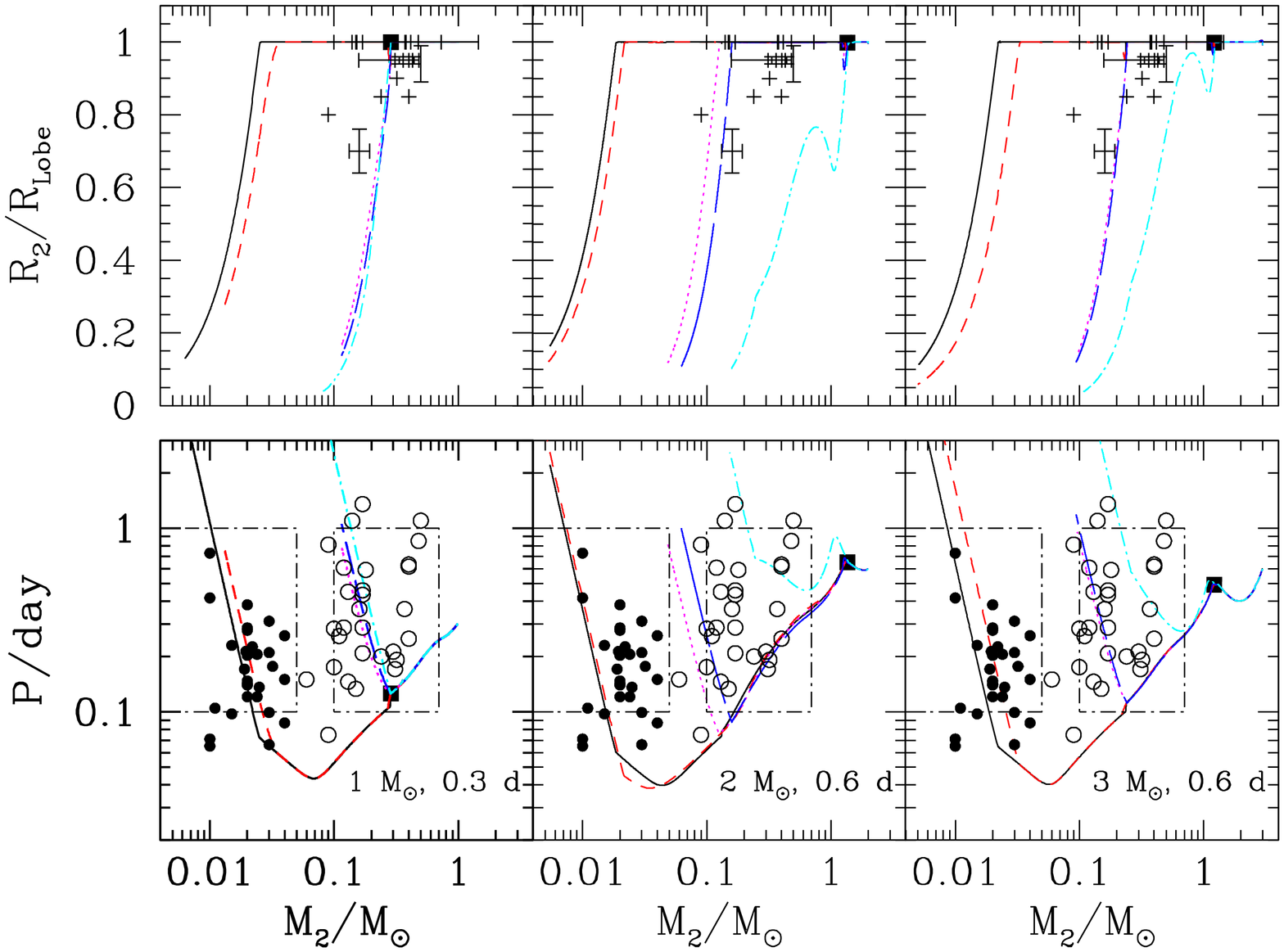}
\caption{The evolution of the same CBSs presented in Fig.~\ref{Fig:figure1}, but
with evaporation starting at the first detachment of the donor due to 
irradiation feedback. Line types, observations, and regions for RBs and BWs are
the same as in Fig.~\ref{Fig:figure1}. Lower panels show the evolution of these
systems, whereas upper panels depict the filling factors. Heavy squares on
the tracks indicate the moment of the first detachment, when evaporation
sets in. While for the case of $M_{2}= 1~M_{\odot}$ detachment occurs
approximately at the same conditions corresponding to the case of models without
irradiation feedback, higher mass models detach when they are far more massive.
Therefore, the entire RBs region  can be populated in this way. As found for
models without irradiation feedback, high evaporation rates lead to low filling
factors, whereas moderate to low evaporation rates lead to (almost) completely
filled lobes.
\label{Fig:figure2}}
\end{figure*}

At this point we consider the possible scenario of irradiation feedback with
radio ejection since first detachment. Let us consider the evolution of the same
CBSs with an intermediate value for $\alpha_{\rm irrad}= 0.10$. This is enough
for our purposes since previous models indicate that the onset of cyclic mass
transfer does not depend strongly on this value (see, e.g. Benvenuto, De
Vito, \&  Horvath 2014; 2015). We assume that  evaporation  starts since the
first detachment of the donor, and consider the same values of $\alpha_{\rm
evap}$ that we have used in the calculations presented in the previous
paragraphs. Models constructed under these hypotheses are presented in
Fig~\ref{Fig:figure2}. The detachment and onset of evaporation of the models
with $1, 2$, and $3~M_{\odot}$ occurs when the donors have masses of  $0.29,
1.36$, and  $1.22~M_{\odot}$ at the ages of $2.10, 1.47$, and $1.10$~Gyr,
respectively. The model of $1~M_{\odot}$ behaves in a way similar to that in
which irradiation feedback is ignored. However, remarkable differences are found
for the cases of the models with $2$ and $3~M_{\odot}$. In these cases
detachment occurs when the models have a mass well {\it above} the high
mass edge of the RB region. Therefore, when they evolve across
it, they may behave as RBs. In this fashion the entire RB region can be
populated. This is in sharp contrast with models that consider evaporation only.
While these models have low filling factors in the case of heavy evaporation,
they also may fill the entire Roche lobe if evaporation is slow.

Finally, we shall refer to the masses of the NSs found in our calculations (see
Table~\ref{tab:masasNSs}). The masses found corresponding to the case of
strongly evaporated models (column 3) are, generally speaking, higher that those
found for the case of irradiation without radio ejection (column 4). On the
other hand, we have made supplementary calculations in which we assumed the same
initial condition (masses and orbital periods) and irradiation feedback, but
without radio ejection. In this case we found slightly higher masses when
comparing them with those of column 3.

\begin{table}
\centering
\caption{The final masses of the NSs for the different conditions considered
in this paper. We give the values for the initial masses of the donor stars and
the orbital periods. In the third column we present the results for the case of
strong evaporation without irradiation. In the fourth and fifth columns we give
the masses for the cases of irradiation feedback with and without radio
ejection, respectively.}
    \begin{tabular}{c|c|c|c|c}
         \hline
         \hline
         $M_{i}/M_{\odot}$ & $P_{i}/$d & $M_{NS}/M_{\odot}$ & $M_{NS}/M_{\odot}$ & $M_{NS}/M_{\odot}$ \\
         \hline
           1 & 0.3 & 1.74 & 1.75 & 1.85 \\
           2 & 0.6 & 2.24 & 1.64 & 2.31 \\
           3 & 0.6 & 2.03 & 1.54 & 2.13 \\
         \hline
         \hline
    \end{tabular}
    \label{tab:masasNSs}
\end{table}

\section{Observational data on redbacks and comparison with theoretical models}
\label{sec:observaciones}

Some RBs have made transitions from accretion to
pulsar state, or vice versa, confirming the model of recycled MSPs in CBSs. This
group is know as transitional millisecond pulsars (tMSPs, PSR~J1023+0038, 
PSR~J1227-4853 and PSR~J1824-2452I). On the other hand, there are some sources 
monitored because they have certain features that make them candidates to
change its status (3FGL~J2039.6-5618, 3FGL~J0838.8-2829, 3FGL~J0212.1+5320,
3FGL~J0954.8-3948, 1FGL~J0523.5-2529,  2FGL~J0846.0+2820).

There are two particularly important states: the RLOF, where the donor star
fills its Roche lobe ($FF=1$), and the quasi-RLOF, where  the donor is slightly 
smaller than its Roche lobe ($FF \approx 1$).

In Table~\ref{table:FF} we present the RBs, or candidates to RBs, known until
today.  In the first column, next to the name of the pulsar and in parentheses,
we label with ``S'' the RBs in Table 1 of \citet{2015MNRAS.446.2540S}; with
``F'' that in the Freire's pulsars catalogue in globular
clusters\footnote{\url{http://www.naic.edu/~pfreire/GCpsr.html}}, but not in
\citet{2015MNRAS.446.2540S}; with ``A'' the systems in the ATNF Pulsar
Catalogue\footnote{\url{http://www.atnf.csiro.au/research/pulsar/psrcat}}
\citep{2005AJ....129.1993M} with $0.1$~d $< P < 1$~d, $0.1 < M_2 / M_{\odot} <
0.7$ and main sequence or unknown companions; with ``P'' those in the Patruno's
catalogue; with ``L'' in \citet{2018MNRAS.473L..50L} and with ``St'' those in
\citet{2019ApJ...872...42S}. In the following columns we list the spin period of
the pulsar, $P_{\rm s}$, the orbital period, $P$, the mass of the
companion, $M_2$ (the best estimation of the mass, or, if it is not available,
the minimum mass\footnote{The minimum mass is computed considering a pulsar mass
of $1.4~M_{\odot}$ and an inclination of $90^o$.}), and the  filling factor $FF$.
As we can see from the Table, most of the RBs do exhibit high filling factor
values. In the cases where $FF$ has an estimation with $FF < 1$, the minimum value
is of $0.70(6)$ for PSR~J1431-4715 \citep{2019ApJ...872...42S} whereas the
maximum value is of $0.92(6)$  for PSR~J1306-40 \citep{2019ApJ...876....8S}.

\begin{table*}

\caption{Relevant observational data for Redback Pulsars. We tabulate the 
name of the pulsar and in parentheses the source of data: ``S'' from
\citet{2015MNRAS.446.2540S}; ``F'' from Freire's catalogue of pulsars in 
globular clusters; ``A'' from the ATNF Pulsar Catalogue 
\citep{2005AJ....129.1993M} with $0.1$~d~$< P < 1$~d, $0.1 < M_2 / M_{\odot}
< 0.7$ and main sequence or unknown companions; ``P'' from Patruno's catalogue;
``L'' from \citet{2018MNRAS.473L..50L} and ``St'' from
\citet{2019ApJ...872...42S}. In the following columns we list the spin period of
the pulsar $P_{s}$, the orbital period $P$, the mass of the companion $M_2$ 
(see main text), and the filling factor $FF$. In the last column we have 
included the main references.}

\begin{threeparttable}
\begin{tabular}{rrrccr}
\hline\hline
PSR & $P_{s}$ [ms] & $P$ [h] & $M_2 [M_{\odot}]$ & $FF$ & Reference \\
\hline\hline
J0024-7204V  (F) & 4.81  & 5.1 & 0.30 - 1.17 & RLOF\tnote{1} & \citet{2016MNRAS.462.2918R} \\ 
J0024-7204W  (P) & 2.35  & 3.2 & 0.15        & RLOF\tnote{1} & \citet{2005ApJ...630.1029B}; \citet{2016MNRAS.462.2918R} \\
J1023+0038   (S) & 1.69  & 4.8 & 0.24        & 0.85\tnote{2} & \citet{2009Sci...324.1411A}; \citet{2015MNRAS.451.3468M} \\
J1048+2339   (A) & 4.66  & 6.0 &  $\sim 0.4$ & 0.85          & \citet{2016ApJ...823..105D}; \citet{2019AA...621L...9Y} \\
J1227-4853   (A) & 1.69  & 6.9 & 0.17 -0.46  & quasi-RLOF\tnote{2} & \citep{2015ApJ...800L..12R} \\
J1306-40     (L) & 2.20  &26.33&  $\sim 0.5$ &     0.92(6)   & \citet{2018MNRAS.473L..50L} \\ 
J1431-4715   (A) & 2.01  &10.8 & 0.13 - 0.19 &     0.70(6)   &\citet{2015MNRAS.446.4019B}; \citet{2019ApJ...872...42S}\tnote{*}\\
J1622-0315   (St)& 3.85  & 3.9 & $\geq 0.10$ & ---           & \citet{2019ApJ...872...42S}\tnote{*} \\
J1628-3205   (S) & 3.21  & 5.0 & 0.17 - 0.24 & quasi-RLOF    & \citet{2014ApJ...795..115L}; \citet{2019ApJ...872...42S}\tnote{*}\\
J1641+3627D  (A) & 3.12  &14.2 & 0.18        & ---           & \citet{2007ApJ...670..363H} \\
J1701-3006B  (S) & 3.59  & 3.5 & 0.13 - 0.41 & RLOF          & \citet{2008ApJ...679L.105C} \\
J1721-1936   (A) &1000.00& 6.2 & 0.11 - 0.27 & quasi-RLOF    & \citet{2005AA...439..433J} \\
J1723-2837   (S) & 1.86  &14.8 & 0.40 - 0.70 & $\sim 1$      & \citet{2013ApJ...776...20C}; \citet{2016ApJ...833L..12V} \\
J1740-5340A  (S) & 3.65  &32.4 & 0.14 - 0.38 & $\sim 0.95$   & \citet{2001ApJ...548L.171D}; \citet{2003AA...397..237O} \\
J1748-2021D  (S) &13.50  & 6.9 & 0.12        & ---           & \citet{2008ApJ...675..670F} \\ 
J1748-2446A  (S) &11.56  & 1.8 & 0.09        & $\sim 0.8$ & \citet{1992ApJ...397..249N}\\
J1748-2446ad (S) & 1.40  &26.3 & 0.14        & RLOF          & \citet{2006cosp...36.1812B}\\
J1748-2446ai (F) &21.23  &20.4 & 0.48        & ---           & Freire's pulsars catalogue \\
J1748-2446P  (S) & 1.73  & 8.7 & 0.38        & RLOF\tnote{1}& \citet{2005Sci...307..892R} \\
J1816+4510   (P) & 3.19  & 8.7 & $\leq 0.19(5)$ & $< 1$ & \citet{2013ApJ...765..158K} \\
J1824-2452H  (S) & 4.62  &10.4 & 0.17        & RLOF          & \citet{2010ApJ...725.1165P} \\
J1824-2452I  (S) & 3.93  &11.0 & 0.17        &  quasi-RLOF\tnote{2}    & \citet{2013Natur.501..517P} \\
J1905+0154A  (A) & 3.19  &19.5 & 0.09        & ---    & \citet{2007ApJ...670..363H} \\ 
J1906+0055   (A) & 2.79  &14.6 & 0.12        & ---         & \citet{2016ApJ...833..192S} \\
J1908+2105   (P) & 2.56  & 3.6 & 0.06        & ---      & \citet{2019ApJ...872...42S}\tnote{*} \\
J1957+2516   (St) & 4.00 & 6.8 & 0.10        & ---      & \citet{2016ApJ...833..192S} \\
J2129-0429   (S) & 7.62  &15.2 & 0.44(4) & $0.95(1)$ & \citet{2016ApJ...816...74B} \\
J2140-2310A  (S) &11.02  & 4.2 & 0.1         &  RLOF?\tnote{3} & \citet{2004ApJ...604..328R} \\
J2215+5135   (S) & 2.61  & 4.1 & 0.33$^{+0.03}_{-0.02}$  & $0.95(1)$ & \citet{2018ApJ...859...54L} \\
J2339-0533   (A) & 2.88  & 4.6 & 0.32        &  $\sim 0.90$ & \citet{2015ApJ...807...18P} \\
\hline\hline
\end{tabular}
   \begin{tablenotes}
     \item[1] From eclipses.
     \item[2] Transitional.
     \item[3] A significant amount of the  material in the eclipsing region is
     outside the companion's Roche lobe.
     \item[*] See also references therein.
   \end{tablenotes}
\end{threeparttable}
\label{table:FF}
\end{table*}

\section{Discussion and Conclusions} \label{sec:conclu}

In this work we have analysed different scenarios for the formation of binary RB
pulsars and its relation with the filled fraction of the Roche lobe associated
to the donor component of the pair. Moreover, we proposed that observations
accurate enough of filling factors may be a powerful tool for discriminating
among the mechanisms of RB formation proposed to date.

Strong evaporation scenarios have been presented by \citet{2013ApJ...775...27C}
and \citet{2015MNRAS.446.2540S}. \citet{2013ApJ...775...27C} assumed the onset
of evaporation since detachment of the donor when it becomes fully convective.
Meanwhile \citet{2015MNRAS.446.2540S} proposed that RBs are due to accretion
induced collapse of a massive WD that detaches the donor from its lobe, allowing
for the onset of evaporation. While close to the onset of  evaporation  the
donor star has a size comparable to its Roche lobe, in most cases soon after
detachment both scenarios lead to low filling factors.

Our models with irradiation feedback (e.g., \citealt{2015ApJ...798...44B}) lead
to high filling factors during the cyclic mass transfer evolutionary stages at
which the system correspond to RB conditions (masses of the components,
evolutionary stage of the donor  star and orbital period). As the NS can
accrete material, in this scenario we expect the occurrence of rather massive
NSs. In this frame, we have analysed the possibility that after the first
detachment induced by irradiation feedback, pulsar irradiation starts out and
inhibits further accretion onto the NS; then, evaporation may drive the systems
to become RBs. If evaporation is strong, it leads to the occurrence of low
filling factors, similar to those predicted by the scenario proposed by
\citet{2013ApJ...775...27C}. However, it is also possible to find binary systems
that evolve throughout the RBs region, essentially with full filled Roche lobes
(i.e., filling factor one). In any case, radio ejection may represent a
difficulty for the existence of some very massive NSs (see, e.g.,
\citealt{2010Natur.467.1081D}, \citealt{2019NatAs.tmp..439C}). 

From the data presented in Table~\ref{table:FF}, there is some indication that
the measured filling factors are generally high ($FF \gtrsim
0.75$). This may, in principle be interpreted as an evidence in favour of
models including irradiation feedback described above. In any case, we should
remark that these non standard evolutionary paths are still  rather uncertain.
These uncertainties are encoded in the free parameters $\alpha_{\rm irr}$ and
$\alpha_{\rm evap}$. We need  a better understanding of the physics of irradiation
and evaporation to get a better scenario of the process of RBs formation. 

We would like to thank our anonymous referee for his/her valuable comments
and suggestions that have helped us to largely improve the original version of
this work.


J.E.H. has been supported by Fapesp (S\~ao Paulo, Brazil) through the grant
2013/26258-4 and CNPq, Brazil funding agencies.


\label{lastpage}

\end{document}